\documentclass[aps,pra,twocolumn,superscriptaddress]{revtex4}

\usepackage{graphicx}
\usepackage{bm}
\usepackage{amssymb,amsmath}
\usepackage{varioref}

\begin{document}

\title{Measurement of diffusion coefficients of francium and rubidium in yttrium based on 
laser spectroscopy}
 
\author{C.~de~Mauro}
\affiliation{CNISM - Unit\`a di Siena and Dipartimento di Fisica, Universit\`a degli Studi di Siena -\\56, via Roma I-53100 Siena (Italy)}
\email{demauro@unisi.it}
\homepage{http://www.unisi.it/fisica}

\author{R.~Calabrese}
\affiliation{Dipartimento di Fisica dell'Universit\`a degli Studi and Istituto Nazionale di Fisica Nucleare, Sezione di Ferrara-\\1, via Saragat I-44100 Ferrara (Italy)}

\author{L.~Corradi} 
\affiliation{Istituto Nazionale di Fisica Nucleare, Laboratori Nazionali di Legnaro -\\2, Viale dell'Universit\`a I-35020 Legnaro (PD) (Italy)}

\author{A.~Dainelli}
\affiliation{Istituto Nazionale di Fisica Nucleare, Laboratori Nazionali di Legnaro -\\2, Viale dell'Universit\`a I-35020 Legnaro (PD) (Italy)}

\author{A.~Khanbekyan} 
\affiliation{CNISM - Unit\`a di Siena and Dipartimento di Fisica, Universit\`a degli Studi di Siena -\\56, via Roma I-53100 Siena (Italy)}

\author{E.~Mariotti} 
\affiliation{CNISM - Unit\`a di Siena and Dipartimento di Fisica, Universit\`a degli Studi di Siena -\\56, via Roma I-53100 Siena (Italy)}

\author{P.~Minguzzi}
\affiliation{CNISM Unit\`a di Pisa and Dipartimento di Fisica, Universit\`a di Pisa\\3, Largo Bruno Pontecorvo I-56127 Pisa (Italy)}

\author{L.~Moi}
\affiliation{CNISM - Unit\`a di Siena and Dipartimento di Fisica, Universit\`a degli Studi di Siena -\\56, via Roma I-53100 Siena (Italy)}

\author{S.~Sanguinetti}
\affiliation{CNISM Unit\`a di Pisa and Dipartimento di Fisica, Universit\`a di Pisa\\3, Largo Bruno Pontecorvo I-56127 Pisa (Italy)}

\author{G.~Stancari}
\affiliation{Dipartimento di Fisica dell'Universit\`a degli Studi and Istituto Nazionale di Fisica Nucleare, Sezione di Ferrara-\\1, via Saragat I-44100 Ferrara (Italy)}

\author{L.~Tomassetti}
\affiliation{Dipartimento di Fisica dell'Universit\`a degli Studi and Istituto Nazionale di Fisica Nucleare, Sezione di Ferrara-\\1, via Saragat I-44100 Ferrara (Italy)}

\author{S.~Veronesi}
\affiliation{CNISM - Unit\`a di Siena and Dipartimento di Fisica, Universit\`a degli Studi di Siena -\\56, via Roma I-53100 Siena (Italy)}

\date{\today}

\begin{abstract}
We report the first measurement of the diffusion coefficients of francium 
and rubidium ions implanted in a yttrium foil. We developed a methodology, 
based on laser spectroscopy, which can be applied to radioactive and stable 
species, and allows us to directly take record of the diffusion time. \\
Francium isotopes are produced via fusion-evaporation
nuclear reaction of a 100~MeV $^{18}$O beam on a Au target at the
Tandem XTU accelerator facility in Legnaro, Italy. Francium is ionized at the
gold-vacuum interface and Fr$^+$ ions are then transported with a
3~keV electrostatic beamline to a cell for neutralization and capture in a magneto-optical trap (MOT). A Rb$^+$ beam is also available,
which follows the same path as Fr$^+$ ions. The accelerated ions
are focused and implanted in a 25~$\mu$m thick yttrium foil for
neutralization: after diffusion to the surface, they are released as neutrals, since the Y work function is lower than the alkali
ionization energies. The time evolution of the MOT and the vapor fluorescence signals are 
used to determine diffusion times of Fr and Rb in Y as a function of temperature. 
\end{abstract}

\maketitle

\section{Introduction}
In many experiments with radioactive atoms, elements are produced as ions and then
neutralized. The most important requirement for a neutralization system is the fast release of neutral atoms, with respect to their radioactive decay time. 
In particular, in our francium experiment at INFN's National 
Laboratories in Legnaro, Italy \cite{lnltrap}, we produce Fr
ions that have to be neutralized before accumulation in a
magneto-optical trap. The system must be very efficient, as a large sample of Fr atoms is requested for
nuclear decay, atomic parity violation and permanent electric
dipole studies.

\begin{figure}[!h]
\centering
\includegraphics[width=\columnwidth]{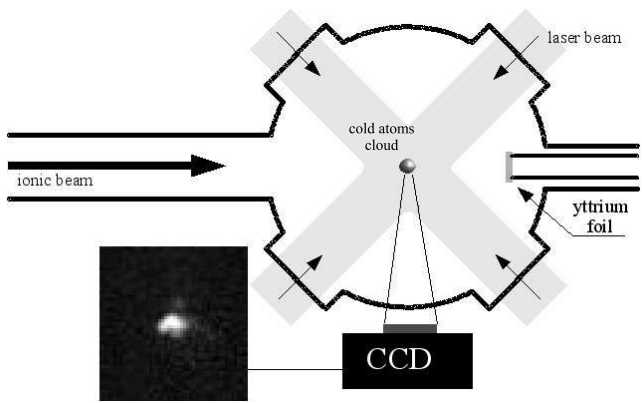}
\caption{Scheme of the experimental setup (the two anti-Helmholtz
coils for magneto optical trapping are not shown). In the inset, an image of the cold Fr trap is reported. The image is not to scale.}
\label{fg:expsetup}
\end{figure}

Our apparatus follows the scheme of the Stony Brook experiment \cite{orozrev}, where many measurements on Fr have been performed so far.
Francium isotopes are produced at the Tandem accelerator facility
in Legnaro via the fusion-evaporation reaction
$$
^{197}\mbox{Au}+^{18}\mbox{O}\longrightarrow
^{215-x}\mbox{Fr}+x\mbox{n}.
$$
With an energy of the primary beam around 100~MeV, we are able to
produce Fr isotopes in the mass number range 208-211. These isotopes 
have a lifetime that is long enough for laser cooling and trapping. The system is at the moment optimized for maximum production of $^{210}$Fr, which has a half-life of
191~s. Francium atoms are ionized at the surface of the gold
target and transported through an electrostatic beamline at an energy
of 3~keV towards a spectroscopic cell designed for laser cooling.
The ionic beam is focused on a 99.96\% pure yttrium foil,
placed inside the cell on the opposite side with respect to the
entrance aperture. In order to be able to detect the trap, we have to optimize all the processes, from production to laser trapping. With only a few days of beamtime per year, it is not convenient to test the setup with francium: we decided to use stable rubidium, always available in high quantities. To this purpose a Rb dispenser was placed in the reaction chamber near the target: Rb atoms that arrive on the gold surface are ionized and injected in the electrostatic line, following the same path as Fr. A sketch of the apparatus is shown in
Fig.~\ref{fg:expsetup}.

Yttrium has a work function $\phi = 3.1$~eV, that is lower than
Fr and Rb ionization potentials, respectively $\mathcal{I}_{\mbox{Fr}}=4.08$~eV and $\mathcal{I}_{\mbox{Rb}}=4.18$~eV. The probability of release in
neutral form at the Y surface is close to unity, according to the
Saha-Langmuir equation
$$
\frac{n_+}{n_a}=\frac{g_+}{g_a}\exp{\left(\frac{\phi-\mathcal{I}}{k_B
T}\right)} ~,
$$
where $n_+$ and $n_a$ are the number of desorbed ions and neutral
atoms, $g_+$ and $g_a$ are the statistical weights due to
degeneracy of the ion and atom ground states respectively
($\frac{g_+}{g_a}=\frac{1}{2}$ for alkalis), and $k_B$ is
Boltzmann's constant. This ratio is $\sim6\times10^{-6}$ at 1000~K.

In spite of the fact that high temperature enhances diffusion, we
fixed an upper limit of 1050~K in order to preserve the cell
coating \cite{wiemots} and hence not to compromise the trapping
efficiency. This is a limitation because melting
point of Y is $T_m$=1799~K, and because the total release
fraction grows with the ratio $T/T_m$ \cite{melconian}.
In this context, our goal was to check whether the efficiency of the 
neutralizer was high enough at a chosen temperature.
So far the release efficiency of radioactive atoms 
embedded in the neutralizer has been measured by monitoring their nuclear decay \cite{melconian}. 
We used atomic laser spectroscopy to observe the time evolution of the 
released neutrals and deduce the diffusion time. We also took advantage of the extraordinary sensitivity 
given by our magneto-optical trap, which allows us to observe as few as 50 
atoms.

\section{1-D Diffusion model}

If an ion is implanted at a given depth inside the Y foil, then it
has to diffuse towards the surface to be released in atomic form.
In order to achieve a good release efficiency, the time necessary
to complete this process has to be small compared with the
radioactive lifetime of the investigated atom. Since the
transverse dimension of the ionic beam is much larger than the
penetration depth, we can use the one-dimensional diffusion
equation for the concentration $N(x,t)$ of the diffusing species
in the Y foil at the distance $x$ from the surface:
$$
\frac{\partial N(x,t)}{\partial t}=D\frac{\partial^2
N(x,t)}{{\partial x}^2}-\Gamma_r N(x,t)+\varphi(x,t) .
$$
$\Gamma_r$ is the
radioactive decay rate ($\Gamma_r=0$ for Rb), $D$ the diffusion
coefficient and $\varphi(x,t)$ is the incoming current of ions impinging on the Y foil. Note that $\varphi(x,t)$ is not directly observable: it 
has to be deduced from the measurement of its integral $I(t)$ (the total 
current) on the neutralizer volume and from the presumed implantation 
distribution, discussed in Sections \ref{sz:pulse} and \ref{sz:cont}.

Empirically, we expect that the dependence of $D$ on the
temperature follows the Arrhenius' law
$$
D=D_{\infty}\exp{\left(-\frac{E_a}{k_B T}\right)}.
$$
$E_a$ is the activation energy and $D_{\infty}$ is the theoretical
asymptotic value of $D$ for large temperatures. The characteristic diffusion time is given by
$$
\tau_d=\frac{d^2}{4D}~,
$$
where $d$ is the mean implantation depth of ions in the Y foil.

\subsection{Pulsed regime} \label{sz:pulse}
In the case of $N_0$ ions all implanted at the same distance $d$
from the surface at the time $t=0$, i.e. $N(x,0) = N_0 \delta(x-d,0)$, the diffusion equation with
$\varphi(x,t)\equiv0$ has an analytical solution: the flux of neutral atoms released in the cell at the time
$t$ is given by~\cite{liatard}
\begin{equation} \label{eq:fl0}
F_0(t)=\frac{2N_0}{\sqrt{\pi}}\sqrt{\frac{\tau_d}{t^3}}\exp{\left(-\frac{\tau_d}{t}\right)}\exp{(-\Gamma_r
t)} .
\end{equation}
For an implantation distribution with a finite extent, the
function can be different, specially for $t<\tau _d$; in a rough
approximation, this difference can be taken into account with a
correction factor, which is expected to be of the order of unity
in the case that $\Gamma_r\ll\tau_d^{-1}$~\cite{liatard}. For
instance, Melconian~\emph{et al.} take the case of a distribution
modelled by a Gaussian times a linear term, with a characteristic
implantation depth $d$~\cite{melconian}. In this case, the
function becomes
\begin{equation} \label{melc}
\tilde{F}_0(t)= \frac{N_0}{2} ~\frac{1}{\tau _d}
~\frac{1}{(1+t/\tau _d )^{3/2}}~ \exp (-\Gamma _r t).
\end{equation}

\subsection{Continuous regime} \label{sz:cont}
If a constant total ionic current $I_0$ is sent to the neutralizer starting at
the time $t=0$, for stable isotopes ($\Gamma_r=0$) we see that the
neutralized released current takes the form
\begin{equation}
F(t)=\int_0^t I_0\;\frac{F_0(t-t')}{N_0}\; dt'=I_0
\left[1-\mbox{erf}\left(\sqrt{\frac{\tau_d}{t}}\right)\right] .
\label{eq:flcw}
\end{equation}

In the case of the implantation distribution considered by
Melconian~\emph{et al.}, we have
\begin{equation} \label{melc2}
\tilde{F}(t)=I_0 \left[ 1- \frac{1}{\sqrt{1+t/\tau _d }} \right] .
\end{equation}

We will see in the following that we can operate in two modes. If
we let ions accumulate in cold yttrium for a certain time and then
we turn on the neutralizer (pulsed mode), all the implanted ions
are released according to Eqs.~(\ref{eq:fl0}) or
(\ref{melc}). Instead, if we turn on the ionic beam with the
neutralizer already working (continuous mode), we are in the case
described by Eqs.~(\ref{eq:flcw}) or (\ref{melc2}).

In our model of the release process, we neglect the contribution of desorption to the release time. We can reasonably assume that, due also to the accuracy of such a model, discussed in Section \ref{sz:contpuls}, desorption contribution is well included in our uncertainty. In any case, as also discussed in Ref. \cite{melconian}, the diffusion coefficients derived from our data analysis can be properly considered as upper limits.

\section{Magneto-optical trapping dynamics}

After being released, atoms can be either directly detected in the vapor
phase (rubidium), or they can be first collected in a small volume (about 1~mm$^3$)
in order to enhance our sensitivity (rubidium and francium). The confinement is provided
by a magneto-optical trap in the standard configuration: six
trapping and repumping laser beams respectively tuned on the D2 and D1 lines,
and a constant gradient magnetic field provided by two anti-Helmholtz
coils.

The total number of atoms in the cell, when the six laser beams
and the magnetic field are present, is the sum of two components:
the number of trapped atoms $N_t$ and the number of atoms in the
vapor phase $N_v$ which is supposed to be at thermal equilibrium
with the cell walls. The time evolution of $N_t$ and $N_v$ is well
described by the coupled rate equations \cite{dema}
\begin{equation} \label{eq:rate}
\left\{\begin{aligned}
\dot{N_t}&=LN_v-CN_t-\Gamma_r N_t\\
\dot{N_v}&=-(L+W)N_v+CN_t-\Gamma_r N_v+f ~,
\end{aligned}\right.
\end{equation}
where $L$ is the rate related to the loading process (the product
$LN_v$ represents the number of atoms per unit time loaded from the
vapor to the MOT), $C$ is the rate describing the collisional
loss of atoms from the MOT ($CN_t$ represents the number of
trapped atoms per unit time which go back from trap to vapor
phase due to collisions with background gas in the cell) and $W$
is the rate characterizing the loss of atoms from the vapor phase
($WN_v$ represents the number of atoms per unit time definitively
lost to the vacuum system or by chemisorption to the cell walls). $f$ is the current of atoms released by the neutralizer and
suitable for trapping. We did not report the quadratic terms due
to collisions between trapped atoms, which are negligible. We can
find the quasi-stationary state solution $\dot{N_t} \simeq
\dot{N_v} \simeq 0$ of Eqs.(\ref{eq:rate}) under the condition
$\dot{f}/f\ll C,W$ (slowly varying incoming current
hypothesis):
\begin{equation} \label{eq:ststsol}
\left\{\begin{aligned}
N_t(t)&=\frac{L}{(W+\Gamma_r)(C+\Gamma_r)+L\Gamma_r}\;f(t) \\
N_v(t)&=\frac{C+\Gamma_r}{(W+\Gamma_r)(C+\Gamma_r)+L\Gamma_r}\;f(t)
\end{aligned}\right.
\end{equation}
In practice, provided that $f$ varies with a time scale much
longer than $1/C$, we obtain that the detected number of trapped
atoms and the total number of atoms in the vapor phase are
proportional to the released current at any time. Therefore, the
measurement of the trap signal or of the fluorescence signal
directly gives all the information about diffusion inside the
neutralizer. This 
constitutes an innovative way of measuring the 
diffusion coefficient, with respect to other experiments that measure the 
release efficiency by detecting the radioactivity from the residual ions in 
the neutralizer.

\section{Measurements methods}

\subsection{Vapor fluorescence and MOT detection}

At room temperature the Rb vapor density is high enough to be directly detected
through resonant excitation and looking at the atomic fluorescence. In order to minimize changes in the apparatus we made measurements by only switching off the trapping magnetic field, but keeping in place the six laser beams. The laser frequency is scanned across the whole profile of the D2 Rb line to search for the maximum signal. Fluorescence light is in this case collected on a Si photodiode. We tried different spectroscopic schemes and we obtained the best signal-to-noise ratio with the one described above.
In the case of francium, it is not possible to directly observe
the atomic vapor fluorescence, because the signals are
extremely weak. The solution is given by the MOT, that allows to accumulate Fr (and also Rb) atoms in a small volume at low temperature. In fact the signal increases by about two orders of magnitude because of Doppler broadening elimination and, at low densities, far from saturated vapor pressures, by several orders of magnitude because of higher densities. As already stressed, we are able to detect traps of only a few tens of atoms. Atomic fluorescence from trapped atoms is detected by a CCD cooled camera (Fig.~\ref{fg:expsetup}). A dedicated software
performs background subtraction, to reduce the contribution of
scattered light to the noise, and real time correction for laser
intensity fluctuations \cite{dema}. The detector was accurately
calibrated and the software directly gives the number of atoms in
the cold cloud.

\subsection{Continuous and pulsed regime} \label{sz:contpuls}
\begin{figure}[!h]
\centering
\includegraphics[width=\columnwidth]{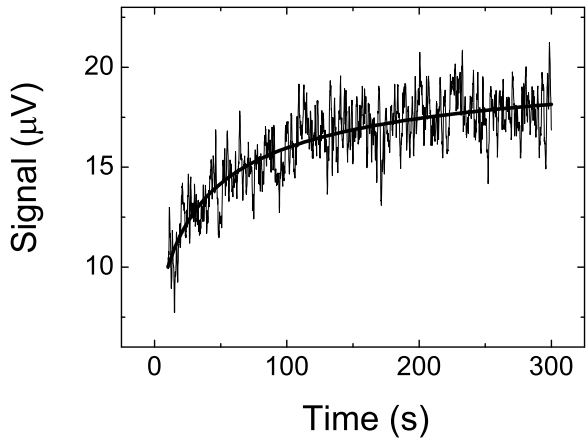}
\caption{Fluorescence signal from the Rb atoms desorbed by the
neutralizer. The curve from Eq.~(\ref{melc2}) (with an additional
offset) was fitted to the experimental data.} \label{contRb}
\end{figure}

When we turn on the ionic beam and send it on a hot neutralizer set at a
constant temperature, we expect the spectroscopic signals to be
well described by Eqs.~(\ref{eq:flcw}) and (\ref{melc2})
(Fig.~\ref{contRb}). The main advantage of this method is that the temperature of 
yttrium is kept constant during the measurement,
and the vacuum conditions for the MOT are very stable. We
analyzed Rb data taken in the continuous regime for several
temperatures. The comparison of the results for the fits with Eq.~(\ref{eq:flcw}) and with
Eq.~(\ref{melc2}) showed an interesting feature, which allows us
to estimate the impact of the implantation distribution on our
results: while the two resulting curves are almost indistinguishable, assuming 
a sharp implantation distribution yields diffusion times that are 
systematically lower by a factor 0.7 with respect to the results obtained in the
hypothesis of an extended initial distribution such as the one
used by Melconian \emph{et al.}. This means that the accuracy for the measured diffusion 
coefficient is anyway limited at the 30\% level by the theoretical model, 
also for previous experiments \cite{melconian,liatard}.

In order to enhance the signals, it is possible to operate in a
pulsed mode. We let the ions accumulate in cold yttrium: at room
temperature, the diffusion coefficient is very small, and the ions stay embedded in the neutralizer. When the neutralizer is
turned on, the atoms are released all together in a time scale of the order of $\tau_d$, according to
Eq.~(\ref{eq:fl0}) or (\ref{melc}). In practice, for both Rb
and Fr we usually kept the ionic current on for 600 s (which corresponds to more than three $^{210}$Fr radioactive half-lives). This
method gave excellent results to estimate the neutralizer release
efficiency, and to determine the best operating temperature for
our neutralizer.

\section{Experimental results}

\subsection{Rubidium} \label{sz:rbmeas}

Most measurements have been done in the pulsed mode,
because of the higher signal-to-noise ratio. We obtained a set of data for each
neutralizer operating temperature: the main goal was to
optimize the yttrium temperature in order to obtain a good release
efficiency.

\begin{figure}[!h]
\centering
\includegraphics[width=\columnwidth]{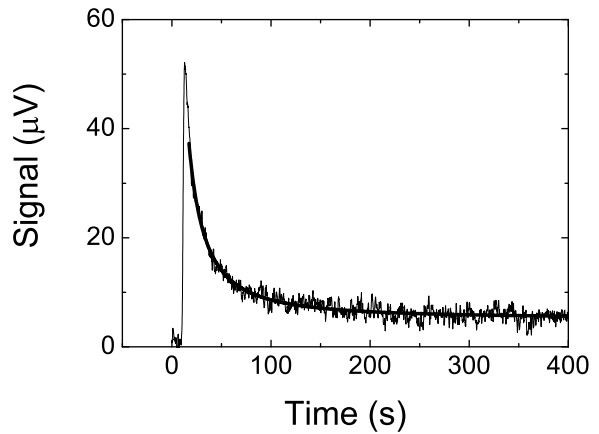}
\caption{Fluorescence signal from Rb
neutralized ions in the pulsed mode and fit curve from
Eq.(\ref{melc}), at a temperature of 960~K.}
\label{fg:tempineutr1}
\end{figure}

The function $G(G_0,S_0,\tau _d
,t_0;t)=G_0+S_0/(1+(t-t_0)/\tau _d )^{3/2}$,
corresponding to the flux $\tilde{F} _0 (t)$ (Eq.~(\ref{melc})),
was fitted to each acquired data set, with the free parameters
$G_0$, $S_0$ and $\tau _d$ (Fig.~\ref{fg:tempineutr1}). The offset $G_0$
is due to the presence of Rb impurities in yttrium: even with
fresh neutralizers, never exposed to Rb current, we detect Rb
vapor in the cell coming from the heated yttrium foil. Note that $t_0$ is not 
left as a free parameter, 
because it is totally correlated with the other parameters $S_0$
and $\tau _d$: for any value $t_0'$, we can find the parameters
$S_0'=S_0 / (1+\frac{t_0'-t_0}{\tau _d})^{3/2}$ and $\tau _d
'=\tau _d ~ (1+\frac{t_0'-t_0}{\tau _d})$, such that
$G(G_0,S_0,\tau _d ,t_0;t) \equiv
G(G_0,S_0',\tau _d ',t_0';t)$. Therefore, $t_0$ is
manually set to the time at which the signal begins rising, just
after turning on the neutralizer. Since the neutralizer does not
reach the operating temperature instantaneously, $t_0$ is affected
by an error corresponding to the signal rising time.

The initial part of the curve critically depends on the way the
neutralizer is turned on, and on the initial distribution of ions
inside yttrium (cf. Eqs.~(\ref{eq:fl0}) and (\ref{melc})).
Therefore, we decided to consider data after a time $t_\text{in}$,
corresponding to a signal which is equal to 3/4 of the maximum
signal for the whole curve (Fig.~\ref{fg:tempineutr1}).

\begin{figure}[!h]
\includegraphics[width=\columnwidth]{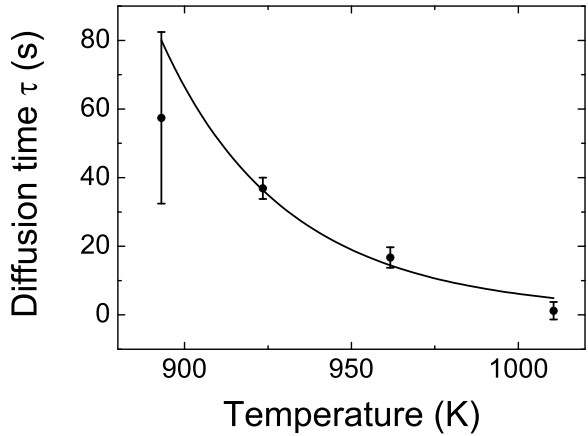}
\caption{Diffusion time of Rb in yttrium as a function of temperature,
fitted with the Arrhenius' function (Eq.(\ref{eq:taud})).}
\label{fg:tempineutr2}%
\end{figure}
\begin{figure}[!h]
\centering
\includegraphics[width=\columnwidth]{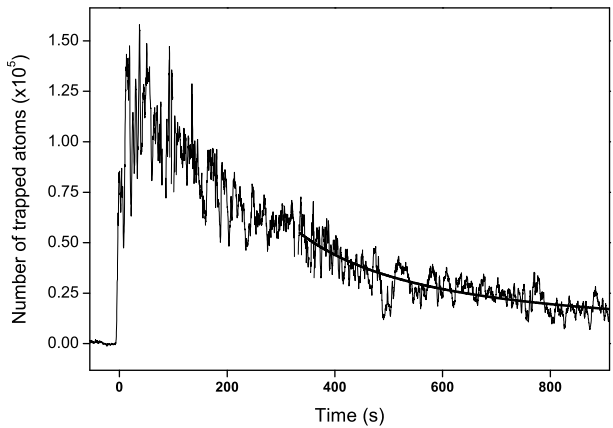}
\caption{Rubidium MOT in pulsed regime: experimental data and fit curve. Neutralizer temperature was about 960~K}
\label{fg:motrb}
\end{figure}

Release time data are consistent with Arrhenius'
law for the diffusion coefficient $D$.
Our experimental results are then fitted accordingly (Fig.~\ref{fg:tempineutr2}):
\begin{equation} \label{eq:taud}
\tau_d=\tau_{1000}\;\exp{\left[\frac{E_a}{k_B}\left(\frac{1}{T}-\frac{1}{1000\mbox{K}}\right)\right]}
\end{equation}
where $\tau_{1000}$ is the diffusion time at 1000~K; fit
parameters are both the time $\tau_{1000}$ and the activation
energy $E_a$. The choice of such a fit function lies in the fact
that $\tau_{1000}$ better describes the diffusion time in the
temperature range of our measurements, while $\tau_\infty$ would
be affected by a larger error, due to the extrapolation to
infinite temperature. Results of fit are $\tau_{1000}=(6.1 \pm
2.3)$~s and $E_a=(1.8 \pm 0.4)$~eV. We used the TRIM code
\cite{trim} to evaluate the mean implantation depth for Rb in Y at
3~keV, which gives a value of 4.9~nm. From this data we can then
deduce the diffusion coefficient at 1000~K:
$$
D_{1000}=\frac{d^2}{4\tau_{1000}}=(1.0\pm 0.4)\times 10^{-14}\;
\mbox{cm}^2\mbox{s}^{-1} .
$$
We also performed measurements on Rb using the MOT detection method, to prepare the Fr 
measurements and to test our diffusion model also for the MOT signal evlution. Experimental data and fit curve
for Rb MOT in pulsed regime are shown in Fig.~(\ref{fg:motrb}). A good agreement was found between data and model, and also
with the results obtained from vapor fluorescence detection described above.
The fit procedure used in this case is discussed in Section~\ref{sz:frmeas}.

\subsection{Francium} \label{sz:frmeas}

The diffusion time of francium in yttrium has been measured using the magneto-optical trap. We
operate in the pulsed mode in order to get higher signals: we use the same method as for Rb, with detection of trapped atoms. With a
collisional rate $C$ of about 0.5~s$^{-1}$, we checked that the
condition $\dot{f}/f \ll C,W$ is verified for all the acquired data.

\begin{figure}[!h]
\centering
\includegraphics[width=\columnwidth]{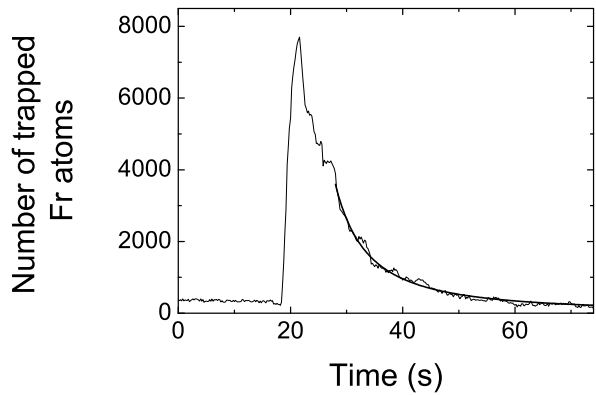}
\caption{Trapped Fr fluorescence and fit curve according to Eq.~(\ref{melc})
with $\Gamma_r=3.633\times10^{-3}\;\mbox{s}^{-1}$.
The exposure time of the CCD camera was set to 1~s.}
\label{fg:trapfit}
\end{figure}

As compared to vapor fluorescence data, MOT fluorescence is quite
difficult to analyze in the pulsed mode, to deduce the diffusion
time. When we turn on the neutralizer, francium atoms come out of
yttrium along with many other impurities already contained in the
neutralizer (as Rb): vacuum suddenly deteriorates, and then slowly
comes back to the equilibrium value. Since the coefficient $C$ in
Eq.~(\ref{eq:ststsol}) strongly depends on vacuum quality, in the
first part of the curve ($t-t_0 < \tau _d$) $C$ depends on time
and it is not easy to interpret the data. However, retaining only data
with $t-t_0 \gg \tau _d$ cannot be a solution, because in this
regime $S_0$ and $\tau _d$ are totally correlated (it is not
possible to distinguish them in the fitting procedure). Our
compromise was to consider data after a time $t_\text{in}$ that
corresponds to a signal which is equal to half the maximum signal
of the whole curve (Fig.~\ref{fg:trapfit}). For these data vacuum
effects are satisfactorily reduced and the order of magnitude is reasonable.

It was not always easy to reach convergence in fits: in practice,
we kept $S_0$ and $\tau _d$ as free parameters, and we set the
offset and $t_0$ to a reasonable value for each fit. In some cases,
$\tau _d $ and $S_0$ were too correlated to obtain a good
convergence: we had to repeat the fitting procedure with $S_0$ set
by hand to realistic values. The error for the obtained parameters includes several
contributions: statistical error from the fit and uncertainties of
$G_0$ and $t_0$.
 
We saw in Section~\ref{sz:rbmeas} that the 
methodology was checked with Rb atoms.
The agreement between MOT and vapor data
demonstrates the validity of our measurements and of the
fitting procedure.

We expect to be able to improve our results. Because the presented
measurements were initially intended to test the final release
efficiency, in terms of number of atoms $N_0$ rather than release
time $\tau _d$, we preferred to begin with the pulsed method,
which gives higher signals and higher signal-to-noise ratios.
Nevertheless, in our limited beam-time we managed to acquire data
in the continuous regime for one neutralizer temperature: we 
remind that in this regime we expect no vacuum-related problems. The analysis for the
obtained curve gave very promising results, in agreement with the
pulsed method.

\begin{figure}[!h]
\centering
\includegraphics[width=\columnwidth]{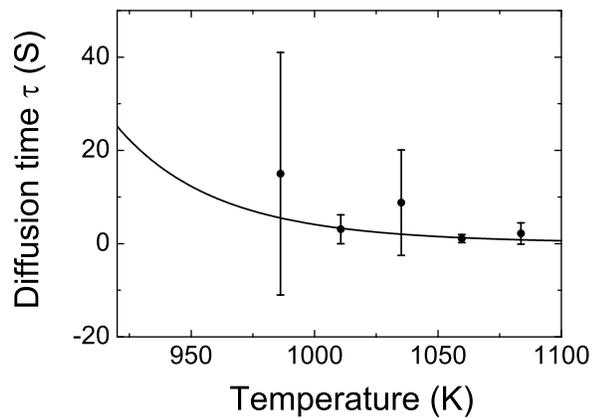}
\caption{Diffusion time of Fr in Y as a function of temperature: experimental data and fit curve  according to Arrhenius' law.} \label{fg:timefit}
\end{figure}

In Fig.~\ref{fg:timefit} the results for the fitted parameter
$\tau_d$ are reported as a function of temperature. We performed release time measurements using also the $\alpha$ decay rate signal, and we obtained similar results, at least as order of magnitude, thus confirming the consistency of our method. Fit results
according to Eq.(\ref{eq:taud}) are:
$$
\tau_{1000}=(4.1 \pm 2.5)\;\mbox{s} \qquad E_a=(1.8 \pm 1.1)\;
\mbox{eV}
$$

An activation energy of $E_a=(1.59 \pm 0.15)$~eV was measured for
$^{37}$K in yttrium in Ref.~\cite{melconian}, so our results for both
Fr and Rb are consistent with literature. We used the TRIM
code \cite{trim} to evaluate the mean implantation depth for Fr in
Y at 3~keV, which gives a value of 5.1~nm. The measured diffusion
coefficient at 1000~K is then
$$
D_{1000}=(1.6 \pm 0.9)\times10^{-14}\;\mbox{cm}^2\mbox{s}^{-1}
$$
This is to our knowledge the first determination of diffusion
time of francium ions in yttrium and the first time that
diffusion parameters are derived from the study of the time evolution
of the signal from a magneto-optical trap.

\begin{figure}[!h]
\centering
\includegraphics[width=\columnwidth]{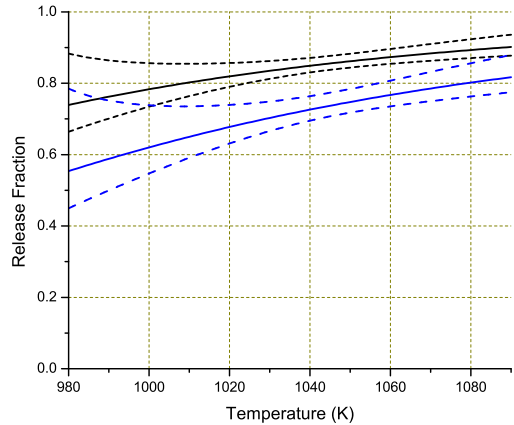}
\caption{(Color online) Calculated release fraction vs temperature for isotope 210 (black curve) and 209 (blue curve), according to Arrhenius' law and $E_a$ and $\tau_{1000}$ given from our fit. Dashed curves give the uncertainty interval.}
\label{fg:relfract}
\end{figure}

According to Ref.~\cite{liatard} it is possible to define a
parameter $\varepsilon
(\Gamma_r)=\exp{(-2\sqrt{\Gamma_r\;\tau_d})}$, which takes into account the loss of atoms due to nuclear decay
during the diffusion towards surface. This is the so-called
``release fraction'' which gives the percentage of radioactive
ions neutralized before the decay and then suitable for trapping.
In other words, one could interpret $\varepsilon
(\Gamma_r)$ as a sort of
neutralization efficiency. From the diffusion time measurement we
determine the release fraction as a function of temperature as
shown in Fig.~\ref{fg:relfract}.

Note that our measurements of the diffusion coefficients for
$^{210}$Fr allow us to predict the release 
fraction for $^{209}$Fr, which we also produce and trap, since we can reasonably suppose that diffusion features are identical for the two isotopes with negligible mass difference.
Because the apparatus was not optimized for the production of this
isotope, we did not perform systematic measurements on $^{209}$Fr.
Nevertheless, our prediction of $\varepsilon$ provides an estimation of the optimal conditions for the $^{209}$Fr
neutralization: due to shorter lifetime of about 72~s, we expect $^{209}$Fr
neutralization to be more efficient at higher temperatures with
respect to $^{210}$Fr~(Fig.~\ref{fg:relfract}). Analog considerations can be applied to $^{211}$Fr: due to very similar radioactive lifetime, the curve calculated for $^{211}$Fr is indistinguishable from the $^{210}$Fr one.

\section{Conclusions}
We measured for the first time the diffusion parameters of francium and 
rubidium ions in yttrium. The methodology that we developed takes advantage 
of laser spectroscopy techniques, and of the very high sensitivity offered 
by laser cooling and trapping. Since we directly measure diffusion times, 
our method can be considered complementary of release efficiency 
measurements by nuclear decay. Another advantage is the possibility to 
measure diffusion coefficients also for stable elements. \\
Experimental data are consistent with the diffusive model
considered, demonstrating the validity and the potential useful
application of our method. The diffusion coefficient is determined
together with an activation energy consistent with previous
measurements. For Fr, a release fraction of more than 80\% is
obtained for the 210 isotope at 1050~K, while for the short-lived 209
isotope we predict a release fraction around 70\% at the same
temperature.


\begin{thebibliography}{7}
\expandafter\ifx\csname natexlab\endcsname\relax\def\natexlab#1{#1}\fi
\expandafter\ifx\csname bibnamefont\endcsname\relax
  \def\bibnamefont#1{#1}\fi
\expandafter\ifx\csname bibfnamefont\endcsname\relax
  \def\bibfnamefont#1{#1}\fi
\expandafter\ifx\csname citenamefont\endcsname\relax
  \def\citenamefont#1{#1}\fi
\expandafter\ifx\csname url\endcsname\relax
  \def\url#1{\texttt{#1}}\fi
\expandafter\ifx\csname urlprefix\endcsname\relax\def\urlprefix{URL }\fi
\providecommand{\bibinfo}[2]{#2}
\providecommand{\eprint}[2][]{\url{#2}}

\bibitem[{\citenamefont{{Atutov} et~al.}(2003)\citenamefont{{Atutov},
  {Calabrese}, {Guidi}, {Mai}, {Scansani}, {Stancari}, {Tomassetti}, {Corradi},
  {Dainelli}, {Biancalana} et~al.}}]{lnltrap}
\bibinfo{author}{\bibfnamefont{S.~N.} \bibnamefont{{Atutov}}},
  \bibinfo{author}{\bibfnamefont{R.}~\bibnamefont{{Calabrese}}},
  \bibinfo{author}{\bibfnamefont{V.}~\bibnamefont{{Guidi}}},
  \bibinfo{author}{\bibfnamefont{B.}~\bibnamefont{{Mai}}},
  \bibinfo{author}{\bibfnamefont{E.}~\bibnamefont{{Scansani}}},
  \bibinfo{author}{\bibfnamefont{G.}~\bibnamefont{{Stancari}}},
  \bibinfo{author}{\bibfnamefont{L.}~\bibnamefont{{Tomassetti}}},
  \bibinfo{author}{\bibfnamefont{L.}~\bibnamefont{{Corradi}}},
  \bibinfo{author}{\bibfnamefont{A.}~\bibnamefont{{Dainelli}}},
  \bibinfo{author}{\bibfnamefont{V.}~\bibnamefont{{Biancalana}}},
  \bibnamefont{et~al.}, \bibinfo{journal}{J. Opt. Soc. Am. B}
  \textbf{\bibinfo{volume}{20}} (\bibinfo{year}{2003}).

\bibitem[{\citenamefont{{Gomez} et~al.}(2006)\citenamefont{{Gomez}, {Orozco},
  and {Sprouse}}}]{orozrev}
\bibinfo{author}{\bibfnamefont{E.}~\bibnamefont{{Gomez}}},
  \bibinfo{author}{\bibfnamefont{L.~A.} \bibnamefont{{Orozco}}},
  \bibnamefont{and} \bibinfo{author}{\bibfnamefont{G.~D.}
  \bibnamefont{{Sprouse}}}, \bibinfo{journal}{Rep. Prog. Phys.}
  \textbf{\bibinfo{volume}{69}}, \bibinfo{pages}{79} (\bibinfo{year}{2006}).

\bibitem[{\citenamefont{{Stephens} et~al.}(1994)\citenamefont{{Stephens},
  {Rhodes}, and {Wieman}}}]{wiemots}
\bibinfo{author}{\bibfnamefont{M.}~\bibnamefont{{Stephens}}},
  \bibinfo{author}{\bibfnamefont{R.}~\bibnamefont{{Rhodes}}}, \bibnamefont{and}
  \bibinfo{author}{\bibfnamefont{C.}~\bibnamefont{{Wieman}}},
  \bibinfo{journal}{J. Appl. Phys.} \textbf{\bibinfo{volume}{76}},
  \bibinfo{pages}{3479} (\bibinfo{year}{1994}).

\bibitem[{\citenamefont{{Melconian} et~al.}(2005)\citenamefont{{Melconian},
  {Trinczek}, {Gorelov}, {Alford}, {Behr}, {D'Auria}, {Dombsky}, {Giesen},
  {Jackson}, {Swanson} et~al.}}]{melconian}
\bibinfo{author}{\bibfnamefont{D.}~\bibnamefont{{Melconian}}},
  \bibinfo{author}{\bibfnamefont{M.}~\bibnamefont{{Trinczek}}},
  \bibinfo{author}{\bibfnamefont{A.}~\bibnamefont{{Gorelov}}},
  \bibinfo{author}{\bibfnamefont{W.~P.} \bibnamefont{{Alford}}},
  \bibinfo{author}{\bibfnamefont{J.~A.} \bibnamefont{{Behr}}},
  \bibinfo{author}{\bibfnamefont{J.~M.} \bibnamefont{{D'Auria}}},
  \bibinfo{author}{\bibfnamefont{M.}~\bibnamefont{{Dombsky}}},
  \bibinfo{author}{\bibfnamefont{U.}~\bibnamefont{{Giesen}}},
  \bibinfo{author}{\bibfnamefont{K.~P.} \bibnamefont{{Jackson}}},
  \bibinfo{author}{\bibfnamefont{T.~B.} \bibnamefont{{Swanson}}},
  \bibnamefont{et~al.}, \bibinfo{journal}{Nucl. Instrum. Methods A}
  \textbf{\bibinfo{volume}{538}}, \bibinfo{pages}{93} (\bibinfo{year}{2005}).

\bibitem[{\citenamefont{{Liatard} et~al.}(1997)\citenamefont{{Liatard},
  {Gimond}, {Perrin}, and {Schussler}}}]{liatard}
\bibinfo{author}{\bibfnamefont{E.}~\bibnamefont{{Liatard}}},
  \bibinfo{author}{\bibfnamefont{G.}~\bibnamefont{{Gimond}}},
  \bibinfo{author}{\bibfnamefont{G.}~\bibnamefont{{Perrin}}}, \bibnamefont{and}
  \bibinfo{author}{\bibfnamefont{F.}~\bibnamefont{{Schussler}}},
  \bibinfo{journal}{Nucl. Instrum. Methods A} \textbf{\bibinfo{volume}{385}},
  \bibinfo{pages}{398} (\bibinfo{year}{1997}).

\bibitem[{\citenamefont{{de Mauro}}(2007)}]{dema}
\bibinfo{author}{\bibfnamefont{C.}~\bibnamefont{{de Mauro}}}, Ph.D. thesis,
  \bibinfo{school}{Universit\`a degli Studi di Siena} (\bibinfo{year}{2007}).

\bibitem[{\citenamefont{{Ziegler}}(2007)}]{trim}
\bibinfo{author}{\bibfnamefont{J.}~\bibnamefont{{Ziegler}}},
  \emph{\bibinfo{title}{\href{http://www.srim.org}{http://www.srim.org}}}
  (\bibinfo{year}{2007}).

\end{thebibliography}
\end{document}